\documentclass[utf8]{frontiersinFPHY_FAMS} 

\setcitestyle{square} 
\usepackage{url,hyperref,lineno,microtype,subcaption}
\usepackage{xspace}
\usepackage{makecell}

\usepackage[onehalfspacing]{setspace}

\newcommand{\pz}{\texttt{p2z}\xspace}
\newcommand{\pr}{\texttt{p2r}\xspace}
\newcommand{\stdpar}{stdpar\xspace}

\newcommand{\mkFit}{\textsc{mkFit}\xspace}
\newcommand{\cpp}{C\texttt{++}\xspace}

\def\keyFont{\fontsize{8}{11}\helveticabold }
\def\firstAuthorLast{Ather {et~al.}} 
\def\Authors{Hammad Ather\,$^{1}$, Sophie Berkman\,$^{2}$, Giuseppe Cerati\,$^{3,*}$, Matti Kortelainen\,$^{3}$, Ka Hei Martin Kwok\,$^{3}$, Steven Lantz\,$^{4}$, Seyong Lee\,$^{5}$, Boyana Norris\,$^{1}$, Michael Reid\,$^{4}$, Allison Reinsvold Hall\,$^{6,*}$, Daniel Riley\,$^{4}$, Alexei Strelchenko\,$^{3}$, Cong Wang\,$^{3,7}$}
\def\Address{$^{1}$University of Oregon, Eugene, OR, USA 97403 \\
$^{2}$Michigan State University, East Lansing, MI, USA 48824 \\
$^{3}$Fermi National Accelerator Laboratory, Batavia, IL, USA 60510\\
$^{4}$Cornell University, Ithaca, NY, USA 14853\\
$^{5}$Oak Ridge National Laboratory, TN, USA 37831\\
$^{6}$United States Naval Academy, Annapolis, MD, USA 21402\\
$^{7}$Clemson University, Clemson, SC, USA 29634
}

\def\corrAuthor{Allison Reinsvold Hall}

\def\corrEmail{United States Naval Academy, Holloway Road,
Annapolis, MD 21402-5002, USA, achall@usna.edu}

\begin{document}

\onecolumn
\firstpage{1}

\title[Code portability solutions for HEP]{Exploring code portability solutions for HEP with a particle tracking test code} 

\author[\firstAuthorLast ]{\Authors} 
\address{} 
\correspondance{} 

\extraAuth{Giuseppe Cerati \\ Fermi National Accelerator Laboratory, PO Box 500, Batavia, IL, USA 60510, cerati@fnal.gov}

\maketitle

\begin{abstract}

Traditionally, high energy physics (HEP) experiments have relied on x86 CPUs for the majority of their significant computing needs. As the field looks ahead to the next generation of experiments such as DUNE and the High-Luminosity LHC, 
the computing demands are expected to increase dramatically. To cope with this increase, it will be necessary to take advantage of all available computing resources, including GPUs from different vendors. 
A broad landscape of code portability tools---including 
compiler pragma-based approaches, abstraction libraries, and 
other tools---allow the same source code to run efficiently on multiple architectures. In this paper, we use a test code taken from a HEP tracking algorithm to compare the performance and experience of implementing different portability solutions.

\tiny
 \keyFont{ \section{Keywords:} heterogeneous computing, portability solutions, heterogeneous architectures, code portability, particle tracking} 
\end{abstract}

\section{Introduction}

Modern high energy physics (HEP) experiments have to process enormous volumes of data in their search to probe extremely rare interactions between fundamental particles. The Compact Muon Solenoid (CMS) experiment~\cite{Chatrchyan:2008aa} at the CERN Large Hadron Collider (LHC), for example, processed hundreds of petabytes of detector data and Monte Carlo (MC) simulations during Run 2 (2015--2018) of the LHC~\cite{CMS_PhaseII_Computing, CMS_PhaseII_ComputingUpdate}. Within the next decade, HEP experiments such as the High-Luminosity LHC (HL-LHC)~\cite{Apollinari:2015wtw} at CERN and the Deep Underground Neutrino Experiment (DUNE)~\cite{DUNE:2020lwj} at Fermilab will pose significant additional computing challenges. The event rate at the LHC is expected to increase by a factor of 7.5, and the data volumes will grow to exabyte scale. Likewise, the expected data rate of a DUNE far-detector module is 9.4 PB per year, and so the total tape volume is expected to exceed the exabyte scale by 2040~\cite{duneCDR}.
To handle these data volumes without sacrificing the physics potential of each experiment, significant R\&D—and accompanying shifts in traditional HEP computing paradigms—are required. 

One paradigm shift that will help HEP experiments prepare for upcoming computing challenges is the ability to utilize parallel heterogeneous computing resources. Historically, the LHC experiments have relied on traditional x86 CPUs for the vast majority of offline computing needs. The majority of the data processing capabilities for the LHC experiments are provided by the Worldwide LHC Computing Grid (WLCG)~\cite{WLCG}, 
which connects 170 computing centers in over 40 countries. 
Increasingly, however, experiments are adapting their software frameworks to take advantage of computing resources at High Performance Computing (HPC) centers~\cite{AtlasCMS_HPC, duneCDR}. All planned exascale platforms rely heavily on GPUs to achieve their anticipated compute performance, and HEP workflows will need to run on GPUs in order to efficiently utilize these resources.

Adapting HEP algorithms to run on GPUs is not a trivial task. For example, CMSSW~\cite{Jones:2006cmssw}, the CMS software framework, includes almost 8 million lines of code~\cite{CMS_PhaseII_Computing}
and was written by hundreds of scientists with varying software backgrounds over the course of decades. Additionally, it is not clear what compute architectures will be prevalent in HPC centers or international scientific computing grids in a decade or two, when the HL-LHC and DUNE experiments are collecting and analyzing data. Even the planned exascale machines in the US use a variety of architectures: Aurora at Argonne National Laboratory uses CPUs and GPUs from Intel, while Frontier at Oak Ridge National Laboratory and El Capitan at Lawrence Livermore Laboratory rely on CPUs and GPUs from AMD. Initial attempts to port HEP algorithms to GPUs typically involved rewriting the original C++ code using CUDA. This process is labor-intensive and only enables offloading to NVIDIA GPUs. Moreover, significant efforts are required to optimize the performance of the initial implementations. The HIP programming language is very similar to CUDA and supports both NVIDIA and AMD GPUS (with very early support for Intel GPUs as well), but neither HIP nor CUDA support CPU architectures directly. Writing and maintaining different implementations for every individual computing platform would take much more expertise and personpower than any HEP experiment can provide.

This is a widely recognized challenge in scientific computing, and there is a broad, rapidly changing landscape of portability solutions that allow a single source code to be compiled and run on a variety of computing backends. The available portability solutions vary widely in terms of overall approach, performance, maturity, and support for different backends or compilers. It is clear, however, that taking advantage of these portability tools will be an essential part of modernizing HEP software. Ideally, a portability solution would achieve two important goals.
First, the portability solution should enable straightforward adaptations of existing HEP algorithms, with minimal rewriting and optimization required. 
Second, the tool should enable algorithms to run efficiently on a variety of different computing architectures, including both CPU and GPU platforms from different manufacturers. 
The performance on different architectures needs to be reasonable, on the same order of magnitude, although it is unlikely to match that of a fully optimized native implementation. 
In this paper, we used a standalone benchmark algorithm to test different code portability solutions and evaluated each in terms of its computational performance and subjective ease of use. 

Programming models and \cpp libraries such as Kokkos~\cite{kokkos-original,kokkos-new} and Alpaka~\cite{alpaka} provide high level data structures and parallel execution options that can be adapted to produce optimized code for a variety of backends, including CPUs and GPUs from NVIDIA, AMD, or Intel (preliminary).  Another portability solution is the \texttt{std::execution::par} (\stdpar) interface, which has been included in the \cpp standard since \cpp17. 
The application programming interface (API) allows for a high level description of concurrent loops, but does not allow for low level optimizations that can be used to enhance performance in native CUDA or HIP.
Various \cpp compilers and associated libraries---such as the oneAPI DP\cpp/\cpp Compiler (dpcpp) from Intel, and nvc++ from NVIDIA---provide support for offloading loops to GPUs, but these compilers are still relatively new.
Similarly, SYCL is a programming model based on the ISO {\cpp}17 standard that enables host and kernel (device) code to be included in the same source file. Finally, there is a category of directive-based portability solutions, which includes OpenMP and OpenACC: through the use of pragmas, developers can specify high level parallelization and memory management behaviors, with the compilers managing the low level optimizations. It should be noted that all these portability solutions, though they are based on open specifications and open-source libraries, generally rely on proprietary, vendor-supplied software stacks (and often compilers) in order to run on particular GPUs.

This paper is organized as follows: In Section~\ref{sec:background}, we describe the motivation and context for this work. In Section~\ref{sec:description}, the benchmark algorithm is described in more detail. The different implementations are covered in Section~\ref{sec:implementations}, including technical details and a subjective discussion of the experience porting the algorithm to each tool. Compute performance results are shown in Section~\ref{sec:results}, and Section~\ref{sec:summary} provides an overall discussion of our experience and lessons learned.

\section{Background and Related Work}
\label{sec:background}

There are several processing steps involved in analyzing data
from a HEP experiment. 
For example, analyzing a time window at the LHC that contains at least one proton-proton collision (referred to as an ``event”), includes the initial data acquisition, ``reconstruction” of the raw detector data into higher level information including  what particles were observed and their energies, and the final data analysis and statistical interpretations. 
This process is similar for other HEP experiments.
In CMS and ATLAS~\cite{ATLAS:2008xda}, the most computationally expensive reconstruction step is track finding, which is the combinatorial process of reconstructing the trajectories of charged particles from the energy deposits (``hits”) they leave in different layers of the detector (see Ref.~\cite{Chatrchyan:2014fea} for a full description from CMS). The standalone benchmarks used for the results in this paper represent the propagation and Kalman update steps (described below) of a traditional Kalman Filter (KF) tracking algorithm~\cite{Fruhwirth:1987fm}. There are two test codes, referred to as the ``propagate to z” or ``propagate to r” benchmarks, denoted by \pz and \pr, respectively. 

The work described in this paper builds off efforts by several other groups working to modernize HEP reconstruction algorithms. The \pz and \pr benchmarks are part of a larger algorithm development effort known as \mkFit~\cite{mkfit_2020}. The goal of the \mkFit project is to rewrite the traditional KF tracking algorithms used by most major HEP experiments and develop a new CPU implementation that is efficient, vectorized, and multithreaded. Depending on the compiler, the \mkFit algorithm achieves up to a factor of six speedup compared to previous KF tracking implementations, and it is now the default algorithm used to reconstruct the majority of tracks in the CMS experiment~\cite{cerati2023chep}. 
The key insight of the \mkFit project is that the KF calculations can be parallelized over the thousands of tracks that may be present within a single detector event. Moreover, if the small matrices and vectors holding the data for each track are arranged in memory so that matching elements from different tracks are stored in adjacent locations, then vector or SIMD  (Single Instruction, Multiple Data) operations can be used to perform the KF calculations.
Similar efforts have also been effective at speeding up code for Liquid Argon Time Projection Chamber (LArTPC) neutrino experiments~\cite{Berkman:2021ffy}.

The \mkFit effort has so far targeted optimizations for Intel multicore CPU architectures such as the Intel Xeon and Intel Xeon Phi processors and coprocessors, but efficient implementations for other architectures will become increasingly important, especially during the HL-LHC era. Given that \mkFit was explicitly designed to create opportunities for vector or SIMD operations, it seems that GPUs should also make a suitable target platform for the \mkFit approach to parallelizing Kalman filtering. 
However, initial attempts to port \mkFit to the NVIDIA Kepler GPU (K40) using CUDA were not very encouraging, both in terms of difficulty and in terms of observed performance (for a full discussion, see Sec. 4 of Ref.~\cite{mkFit_CTD2017}). 
The irregular patterns of memory access that are occasionally needed in order to reorganize the data coming from different tracks turned out to be particularly challenging to manage on GPUs. Even with well-structured data, however, translating standard C++ code to be compatible with NVIDIA CUDA required significant low level re-coding effort to achieve acceptable performance for the basic KF operations. Since it is not feasible to rewrite \mkFit for every possible architecture, the \pz project was started to explore code portability tools in the context of charged particle tracking.

A broader effort with similar motivation is the HEP Computational Center for Excellence (HEP-CCE) collaboration’s Portable Parallelization Strategies (PPS) activity~\cite{ccepps1, ccepps2}. The HEP-CCE PPS project is exploring portability solutions using representative reconstruction algorithms from CMS and ATLAS as well as LArTPC neutrino experiments such as DUNE. 
Collaborators from the HEP-CCE PPS activity became involved in this project and have used the \pr benchmark as a compact, standalone application (mini app) to evaluate GPU offloading via different technologies, such as CUDA, OpenACC, and \stdpar, as described in detail below.

Having two teams working simultaneously on these two complementary mini apps has proven to be important to the project’s success. For many of the implementations described in Section~\ref{sec:implementations}, we found that it was relatively straightforward to do an initial porting of the algorithm but fairly difficult to have a fully optimized version. Different initial strategies in porting the \pz or \pr benchmark meant that multiple approaches could be simultaneously developed and tested. In several cases, an issue was identified in a specific \pz or \pr implementation and the solution was propagated to both mini apps. Having two different teams with unique expertise also expanded the number of portability technologies we could test. Finally, as explained in Section~\ref{sec:results}, slightly different approaches were taken to measure the final results, giving additional insight into  the performance of each tool.

\section{Description of algorithm}
\label{sec:description}

Track finding (also known as track building) is the process of reconstructing a particle’s trajectory by identifying which hits in an event likely came from the same particle. It requires testing many potential combinations of hits to find a set that is consistent with the expected helical trajectory of a charged particle in a magnetic field. Track fitting, on the other hand, is the process of taking a pre-determined set of hits and determining the final parameters of the track. The \pz and \pr benchmarks include everything that would be needed for a realistic track \emph{fitting} algorithm, but do not include the combinatorial selection required for track finding.

The CMS and ATLAS detectors are divided into two main sections: the cylindrical region coaxial with the beam pipe, known as the barrel, and a disk region on either end of the barrel, known as the endcaps. To first approximation, the individual tracker layers can be approximated as being located at constant radius $r$ or constant $z$ position for the barrel and endcap layers, respectively. Charged particles in a constant magnetic field will travel in a helix, so if the position and momentum are known on layer $N$, then the expected position can be calculated for layer $N+1$. 

The mini apps used in this analysis perform two key steps of KF tracking:

\begin{enumerate}
    \item \textbf{Track propagation:} Propagate the track state---including the track’s momentum and position vectors and associated uncertainties in the form of a covariance matrix---at
    layer $N$ to a prediction at layer $N+1$, which is specified by either a $z$ coordinate or a radius $r$ for the \pz and \pr benchmarks, respectively.
    \item \textbf{Kalman update:} Update the track state on layer $N+1$ by combining information about the propagated track state and the coordinates of a compatible hit on that layer. 
\end{enumerate}

These two steps are the most arithmetically intensive steps of both track finding and track fitting. They are relatively simple algorithms but have the biggest impact on the overall execution time, making them suitable for a standalone test code.

\subsection{Input data}

The starting point for a KF track finding algorithm is a track ``seed”, an initial guess at the track state.
For the full \mkFit algorithm, the input track seeds are built in an upstream algorithm using three or four hits from the innermost layers of the CMS detector. For simplicity, the \pz and \pr benchmarks use an artificial standalone input consisting of a single track seed. The parameters of the initial seed are smeared according to a Gaussian distribution, in order to prevent the algorithm from performing identical numerical operations for each track.

The tracks are built by first propagating the initial seed parameters to the next layer. In the full combinatorial \mkFit application, the next step is to search for compatible hits on that layer. In the \pz and \pr benchmarks, only one hit per layer is considered, similar to what is required for track fitting.
The hit parameters---the hit locations and uncertainties on each layer---are smeared per track
according to the same procedure as the input track parameters. 

For the GPU implementations of the benchmark, the track propagation and Kalman update steps are run as a single GPU kernel. The data for all tracks and hits are prepared on the CPU, transferred to the GPU for the computations in two bulk transfers, and the output data are transferred back to the CPU.

\subsection{Computations and data structure}

The ``propagation to z” or \pz benchmark uses the expected helical trajectory of the charged particle to calculate the track parameters and the covariance matrix on endcap layer $N+1$. In contrast, the ``propagation to r” or \pr benchmark uses an iterative approach to propagation, advancing the track state from the initial barrel radius to the final radius in discrete steps. In practice, both approaches involve a series of operations involving $\sin()$ and $\cos()$ functions, as well as matrix multiplication of up to 6x6 matrices.  The two benchmarks are expected to be similar in terms of arithmetic intensity.

The second task in both benchmarks is the Kalman update step, which ``updates” the track state using the parameters of a hit on that layer. The hit parameters include three coordinates for the hit position and a 3x3 covariance matrix. Similar to the propagation step, the update step involves small matrix multiplication and matrix inverse operations as well as trigonometric functions.

\begin{figure}[h!]
\begin{center}
\includegraphics[width=10cm]{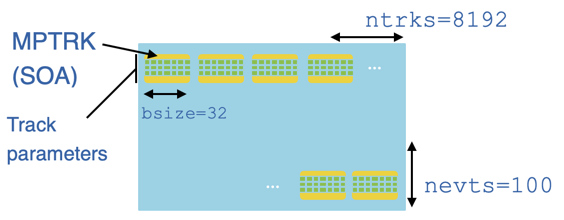}
\end{center}
\caption{Representation of the data structure used in the \pr benchmark. The \pz data structure is similar, but has \texttt{ntrks} equal to 9600.}
\label{fig:dataStruct}
\end{figure}

Both benchmarks employ an Array-Of-Structures-Of-Arrays (AOSOA) data structure, as illustrated in Figure~\ref{fig:dataStruct}. Each benchmark runs over a fixed number of events (\texttt{nevts}) and processes a fixed number of tracks in each event (\texttt{ntrks}). Within one event, tracks are grouped into batches of \texttt{bsize} tracks, and each batch of tracks is put into a Structure-Of-Arrays construct referred to as an \texttt{MPTRK}. Here, as in the full \mkFit algorithm, the goal of organizing tracks into different batches is to enable SIMD operations across batch elements. The value of \texttt{bsize} can be optimized for different platforms; for example, on a GPU it might be the NVIDIA warp size of 32, while on a CPU it might be a multiple of the AVX-512 vector width of 16; for consistency we use 32 everywhere. 

To maximize the opportunities for SIMD operations, the storage order in the AOSOA follows the same general scheme as in \mkFit. Within an \texttt{MPTRK}, the first elements of \texttt{bsize} different vectors (or matrices) get stored in adjacent locations, followed by the second elements of \texttt{bsize} different vectors (or matrices), and so on, until the full structure representing a batch of \texttt{bsize} tracks is completely populated. Then the next \texttt{MPTRK} structure is written into memory, and the next, until all \texttt{ntrks} tracks (8192 for \pr and 9600 for \pz) for the first event are present in memory. This first event corresponds to the first row of SOAs in Figure~\ref{fig:dataStruct}. It is followed by a second row of SOAs for the second event, etc. A similar memory layout applies to the hit data.

\section{Implementations}
\label{sec:implementations}


The state-of-the-art of portability tools is a moving target, as many tools are undergoing very active development, with additional features, backends, and compiler support being added on a regular basis ($\mathcal{O}$(monthly)). In total, we tested nine different parallelization tools on four different architectures,
but testing the full matrix of possibilities was beyond the scope of this paper. The final set of \pr implementations is shown in Table~\ref{table:p2r}, including which backends and compilers were used to test each implementation, and the full set of \pz implementations is shown in Table~\ref{table:p2z}. Note that this does not include the full set of backends or compilers that each tool is capable of supporting.

For both test codes, the original CPU implementation was multithreaded using the oneAPI Threading Building Blocks (TBB) library and compiled with gcc,
since this is the combination that most closely matches what the \mkFit project uses for its highly optimized implementation.
 The reference GPU implementation is the one written using CUDA and compiled with nvcc. 
Below we describe each tool, its corresponding \pr and \pz implementations,  and our subjective experience porting these benchmarks using the different portability solutions. 

\hfill

\begin{table}[h!]
    \centering
\begin{tabular}{|c|c|c|c|c| }
    \hline
& \textbf{NVIDIA GPU} & \textbf{AMD GPU} & \textbf{Intel GPU} & \textbf{x86 CPU} \\

\hline

TBB (oneTBB/2021.10.0) & - & - & - &  gcc/12.3.0\\

\hline

CUDA & \makecell{cuda/11.6.2\\nvcc/11.6.124} & - & - & - \\

\hline

HIP & - & rocm/5.2.0 & - & - \\

\hline

Alpaka (v0.9.0) & \makecell{cuda/11.6.2\\nvcc/11.6.124} & \makecell{ rocm/5.1.3\\hipcc/3.5.0} & -& \makecell{oneTBB/2021.10.0 \\ gcc/12.3.0}  \\

\hline

Kokkos (v3.6.1) & \makecell{cuda/11.6.2\\nvcc/11.6.124} & \makecell{ rocm/5.1.3\\hipcc/3.5.0} &  \makecell{ Kokkos 4.0\\dpcpp/2023.0.0} & \makecell{Kokkos 4.0.0\\gcc/12.3.0} \\

\hline

SYCL & \makecell{cuda/11.6.2\\intel/llvm-sycl} & \makecell{rocm/5.1.3\\intel/llvm-sycl} &  dpcpp/2023.0.0 &  dpcpp/2023.1.0 \\

\hline

\stdpar & nvc++/22.7 & - & \makecell{dpcpp/2023.0.0\\dpl/2022.0.0} & nvc++/24.5 \\

    \hline        
    
    \end{tabular}
    
    \caption{Summary of \pr implementations used for the results shown in Section~\ref{sec:results}. The table lists the     
    compiler versions used for each implementation. For SYCL, the intel/llvm-sycl repository is used to compile dpc++ tool chains to be compatible with NVIDIA/AMD GPUs. 
    This is not the full list of all possible combinations of tools and backends. For example, HIP could be compiled for NVIDIA GPUs as well, but that was not tested in this work. 
    Similarly, we note that the oneAPI toolkit recently included a plugin that enables \stdpar to be used on AMD GPUs, and Alpaka has recently introduced experimental support for Intel GPUs in v0.9.0, but neither was tested here.}
    
    \label{table:p2r}
\end{table}

\hfill

\begin{table}[h!]
    \centering
    \begin{tabular}{|c|c|c| }
    \hline
& \textbf{NVIDIA GPU} & \textbf{x86 CPU} \\

\hline

TBB (oneTBB/2021.10.0) & - & gcc/12.3.0\\

\hline

CUDA & \makecell{cuda/11.2\\nvcc/11.0.221} & - \\

\hline

Alpaka (v0.8.0) & \makecell{cuda/11.2\\nvcc/11.0.221} &  \makecell{Alpaka v0.9.0 \\oneTBB/2021.10.20 \\ gcc/12.3.0}  \\

\hline

Kokkos (v4.0) & \makecell{cuda/11.2\\nvcc/11.0.221} & gcc/12.3.0 \\

\hline

\stdpar & nvc++/24.5 & nvc++/24.5 \\

\hline    

OpenMPv4 & \makecell{OpenARC/0.76~\cite{openarc_hpdc14}\\nvcc/11.0.221} & gcc/12.3.0 \\

\hline

OpenACC & \makecell{OpenARC/0.76~\cite{openarc_hpdc14}\\nvcc/11.0.221} & - \\

\hline
    
    \end{tabular}
\caption{
Summary of \pz implementations used for the results shown in Section~\ref{sec:results}. The table lists the     
    compiler versions used for each implementation.}
    \label{table:p2z}
\end{table}

\hfill

\subsection{TBB }

Our reference CPU implementation is based on the oneAPI Threading Building Blocks (or oneTBB, or just TBB) library. TBB is a template library originated by Intel for parallel programming on multi-core processors that simplifies the definition and optimization of threaded applications. For a given executable, TBB collects all tasks that are specified to run in parallel and the library manages and schedules threads to execute them. TBB is used as the CPU thread manager in the software framework of the CMS experiment.

In our code, nested TBB \texttt{parallel\_for} loops are over events and bunches of tracks. Each bunch of tracks is then vectorized, so that tracks in the bunch are processed in a SIMD fashion. Vectorization is implemented following the 
approach used in \mkFit, where groups of matrices are processed concurrently and loops over the matrix index in the group are decorated with \texttt{omp simd} pragmas. These pragmas were activated by the \texttt{-fopenmp} compiler option.

\subsection{CUDA and HIP}
\label{sec:cudaandhip}

Our reference GPU implementation is based on the CUDA programming model, which is a multi-threaded SIMD model for general purpose GPU programming, introduced by NVIDIA.
The CUDA implementation is ported from the TBB version, which shares the same AOSOA data structure for input data. The main difference in CUDA is that each \texttt{MPTRK} is processed by a block of GPU threads, and each thread processes the computation for one track in the \texttt{MPTRK}. Since the computation of each track is independent of the others, we find keeping the intermediate results in the local registers to have the most efficient memory access. We explored using shared memory to store the intermediate results within an \texttt{MPTRK} for all the threads in the block, but it was shown to have significantly lower memory throughput in a detailed profiling study.

In relation to other portability technologies, CUDA provides a level of abstraction similar to the general accelerator execution model and memory model that is also employed by general GPU programming models such as OpenCL and SYCL. As a proprietary NVIDIA GPU programming model, however, it exposes several NVIDIA-GPU-specific features, which allows available architecture-specific features to be fully exploited, but it also means that the code is not portable across heterogeneous accelerators.

HIP is the vendor-supported, native programming model for AMD GPUs and is designed to be portable across NVIDIA and AMD GPUs. 
It is also designed to be syntactically similar to CUDA so that most API calls can be simple translations of names.
In the case of \pz and \pr, the kernels only rely on the core functionalities of CUDA, such as memory allocation and kernel dispatches, which are supported in HIP.
This leads to a straightforward port to the HIP version starting from the CUDA version.

\subsection{Directive-based solutions, OpenMP and OpenACC }

Directive-based, high-level programming models such as OpenMP and OpenACC use a set of directives (a  special type of comments that a compiler can understand) that allow a programmer to provide the compilers with important characteristics of an application, such as available parallelism and data sharing/mapping rules, so that much of the low-level programming and optimization burdens are automatically handled by the directive compilers.

The biggest advantage of the directive programming model is that it allows incremental porting of existing sequential CPU applications written in standard programming languages such as C, C++, and Fortran to parallel versions that can offload work to heterogeneous accelerators, without requiring major changes in the existing code structures.
The initial OpenMP version was created by converting the reference TBB CPU implementation into an OpenMP CPU implementation, which is relatively straightforward due to the similarity between the TBB parallel syntax (using lambdas) and OpenMP parallel constructs. 
Then, the GPU offloading version was created by extending the OpenMP CPU implementation with OpenMP target-offloading directives and data mapping directives. 
The CPU version and the GPU version have different parallelism and data mapping strategies.
For instance, on GPUs, the OpenMP target and data mapping directives are essential, but when targeting CPUs, they are unnecessary since the original host data and corresponding device data will share storage; in the latter case, how the OpenMP compiler implements those unnecessary directives on the CPUs is implementation-defined.
Another issue is that on CPUs, team-level parallelism may or may not be ignored depending on the compiler implementations, which may incur additional overheads due to inefficient mapping.

The initial conversion from the OpenMP GPU implementation into the OpenACC implementation was straightforward since both OpenMP and OpenACC 
are directive-based accelerator programming models and provide very similar execution and memory models.
However, the main issue in converting between OpenMP and OpenACC was that different OpenMP/OpenACC compilers may choose different parallelism mapping strategies and vary in terms of maturity and supported features (see  Section~\ref{sec:compilers} for results and a more detailed discussion).

\subsection{Alpaka}

Alpaka~\cite{alpaka} is a single-source, header-only \cpp parallelization library. 
The API level is similar to CUDA, with an abstraction layer added between the application and the vendor-specific programming models to achieve portability.
For example, the kernel functions in Alpaka are templated with an accelerator type, which is resolved at compile time for different execution backends of the kernel. 
One difference between Alpaka and CUDA is that Alpaka has an additional abstraction level called \textit{elements} in the parallel hierarchy model, where multiple elements can be processed in a thread.
Having an explicit level allows compilers to take advantage of the SIMD vector registers when compiling for the CPU backends.
Each SOA (\texttt{MPTRK}) is processed by a block, and the thread/element level is mapped differently between CPU and GPU backends to take full advantage of the parallel hierarchy.
For GPU backends, blocks of \texttt{bsize} threads are assigned to process each \texttt{MPTRK}, whereas one thread with \texttt{bsize} elements is assigned to process each \texttt{MPTRK} for CPU backends. In each case, enough threads are defined so that all \texttt{ntrks} tracks are processed for an event.
Profiling results confirm the use of vectorized instructions in the regions where the original \mkFit implementation was also able to vectorize. 

The overall conversion from CUDA to Alpaka is relatively smooth due to the similarity between the two programming models, except that the heavy use of templating often leads to a more verbose coding style and  convoluted error messages during debugging.
Nevertheless, Alpaka versions are often able to produce close-to-native performance after some effort of optimization. 
A particular relevant note for HEP experiment is that CMS has chosen to use Alpaka as its supported portability solution for the GPU usage in LHC Run 3~\cite{kortelainen:2021a,bocci:2023}.

\subsection{Kokkos}
Similar to Alpaka, Kokkos~\cite{kokkos-original,kokkos-new} serves as a single-source \cpp template metaprogramming (TMP) library, intended to achieve architecture agnosticism and alleviate programmers from the complexities of vendor- or target-specific programming paradigms and heterogeneous hardware architectures. 
By embracing TMP methodologies, Kokkos facilitates device-specific code generation and optimizations via template specialization. 
To cater to diverse computing environments, Kokkos offers multiple device-specific backends, implemented as template libraries atop various HPC programming models like CUDA, HIP, OpenMP, HPX, SYCL, and OpenACC. 
These backends are tailored to adhere to advancements in the \cpp standard, ensuring compatibility and efficacy. 

A notable departure from Alpaka is Kokkos' emphasis on descriptive rather than prescriptive parallelism. 
Kokkos prompts developers to articulate algorithms in general parallel programming concepts, which are subsequently mapped to hardware by the Kokkos framework. 
The Kokkos programming model revolves around two fundamental abstractions: the first being the user data abstraction (\verb|Kokkos::View|), a template library facilitating the representation of multidimensional arrays while  managing efficient data layout for both CPU and GPU. 
The second abstraction revolves around parallel execution patterns (\verb|parallel_for|, \verb|parallel_reduce|, and \verb|parallel_scan|), which can be executed under three distinct execution policies: \verb|RangePolicy| for mapping single parallel loops, \verb|MDRangePolicy| for mapping directly nested parallel loops, and \verb|TeamPolicy| for hierarchical mapping of multiple nested parallel loops. 

In this study, Kokkos versions were developed by translating CUDA code to Kokkos using the Kokkos \verb|View| and parallel dispatch abstractions. 
The initial translation process was relatively straightforward, owing to the striking similarities between the Kokkos and CUDA execution models and memory models, except for the added complexity due to Kokkos-specific restrictions on \cpp template programming. 
Efficient execution on both CPU and GPU took further optimization and was
achieved by  configuring the \verb|teamSize| and \verb|vectorSize| in the \verb|TeamPolicy| used in the \verb|parallel_for| execution pattern, as the defaults were found not to be optimal.

\subsection{Standard parallelization using \stdpar in \cpp }


The C++ programming language is often the preferred choice for implementing high performance scientific applications. The recent revisions of the ISO C++ standard
introduced a suite of algorithms capable of being executed on accelerators. Although this approach may not yield the best performance, it can present a viable balance between code productivity and computational efficiency.
Numerous production-grade compilers are available, 
such as clang and its variants from various providers, or the recently released NVHPC from NVIDIA.

With the introduction of the
C++17 standard, the \cpp Standard Template Library (STL) underwent a substantial overhaul of its suite of
algorithms, now updated with execution policies to adapt across various computing architectures,
including multi-core x86 systems and GPUs. These parallel algorithms extended most of the existing STL algorithms with an additional argument, which is an execution policy. The policy enables programmers to specify the intended parallelism of an algorithm, which can result in performance improvements for the computational task.
In particular, the execution policies in C++17 include:
\begin{itemize}
\setlength\itemsep{-0.5em}
\item \texttt{std::execution::seq}
\item \texttt{std::execution::unseq}
\item \texttt{std::execution::par}
\item \texttt{std::execution::par\_unseq}
\end{itemize}
The first option forces the algorithm to run sequentially, while the remaining three options allow the algorithm to be vectorized or run in parallel (with additional vectorization). 
Both the \pz and \pr \stdpar implementations use the \texttt{std::execution::par\_unseq} execution policy.
Currently, only the nvc++ compiler offers support for \stdpar algorithms to be offloaded on NVIDIA GPUs. It leverages CUDA Unified Memory to handle automatic data movement between CPU and GPU. On the systems that do not support Heterogeneous Memory Management, only data that is dynamically allocated in CPU code compiled by nvc++ is automatically managed, whereas memory allocated in GPU code is exclusively for GPU use and remains unmanaged. Thus, on such systems, CPU and GPU stack memory, along with global objects, are outside nvc++'s automatic management scope. Even data allocated on the CPU heap outside units compiled by nvc++ is not managed. When dealing with Parallel Algorithm invocations, pointers and objects must refer to data on the managed CPU heap to avoid errors. Any dereferencing of pointers to CPU stack or global objects in GPU code can lead to memory violations. These aspects encapsulate the nvc++ compiler's precise approach to memory management across CPU and GPU, emphasizing careful allocation and reference handling to ensure efficient operations. In our experience, developing code for this application is largely similar to standard C++ programming, with the primary distinction being the need to consider the previously mentioned limitations.


\subsection{SYCL }

SYCL represents a cross-platform abstraction layer that enables code for heterogeneous processors to be written in a “single-source” style using completely standard C++. This approach aims to enhance the efficiency and accessibility of programming for a variety of compute architectures.
While the SYCL programming language was promoted by the Khronos Group, it is predominantly advocated for by Intel, so the primary focus is on the optimization for Intel GPUs. 

One of the key advantages of SYCL is the ability to handle regular C++ code for the host CPU and a subset of C++ for the device code within the same source file. This ability paves the way for an integrated and simplified development process. It also enables parallelism and the usage of memory hierarchies through a class template library. This effectively allows the expression of parallel STL, thus further integrating SYCL with standard C++ features. Since it was designed to be fully compatible with standard C++, it allows developers to utilize any C++ library within a SYCL application. This compatibility with standard C++ makes SYCL a versatile tool for developers. Moreover, SYCL's design prioritizes performance portability, aiming to provide high performance across a wide range of hardware architectures. Its abstractions are constructed to allow optimization but do not require a particular architecture or kernel language.

For the benchmarks described in this paper, the programming approach for SYCL is nearly a direct replication of the CUDA approach. In our implementation, we utilized SYCL's Unified Shared Memory feature for data management. To compile SYCL for NVIDIA and AMD GPUs, we compiled the dpc++ tool chain following the instructions in Intel's open-source llvm-sycl repository\footnote{Intel llvm/sycl branch, https://github.com/intel/llvm/tree/70c2dc6dcf73f645248aa7c70c8cefdabf37e9b7}.

\section{Results}
\label{sec:results}

The most important computing metric, from the point of view of  HEP computing as a whole, is the algorithm’s throughput, defined here as the number of tracks that can be processed per second. 
To measure the throughput, 
all versions processed approximately 800k tracks, with the same settings for number of iterations and number of iterations per track. The measurements were repeated multiple times to ensure stability of results. We tested the performance of different tools on a number of different hardware systems, including NVIDIA GPUs (Section~\ref{sec:NVIDIA}), Intel and AMD GPUs (Section~\ref{sec:AMD_Intel}), and Intel CPUs (Section~\ref{sec:CPU}).

The code used for the results in this paper is directly extracted from the \mkFit application. In order to reduce overheads, it does not have its own setup for validating the results in terms of physics output. For a discussion of the \mkFit physics performance results, see the \mkFit papers~\cite{mkFit_CTD2017,mkfit_2020}.
Numerical reproducibility and consistency of results across different \pz and \pr versions are verified with summary printouts. Since the artificial sample is produced by smearing the inputs from a single track, printouts report the mean and RMS values of the track parameters computed over the full sample.
Output results from different versions are numerically identical when the same compiler is used. Compilers may introduce differences in the average parameter values at or below the ppm level. Such differences are due to numerical precision in floating point operations and the different levels of optimization used by default by different compilers. The level of agreement in the output guarantees that all versions perform the same operations, making the comparisons in terms of computing performance meaningful.




\subsection{NVIDIA GPU results} 
\label{sec:NVIDIA}


The throughput of the \pz and \pr benchmarks were measured using two different systems with two different NVIDIA GPUs. 
To test the \pr implementations, we used the Joint Laboratory for System Evaluation (JLSE), a collection of HPC testbeds hosted at Argonne National Lab. The NVIDIA GPU that was used for testing is an A100 GPU with an AMD 7532 CPU as the host machine.
For the \pz implementations, the measurements were performed on a test node for the Summit HPC system at Oak Ridge National Laboratory. Each Summit node includes six NVIDIA V100 GPUs and two IBM Power9 CPUs, although only one GPU was utilized in our tests.

The throughput measurements on NVIDIA GPUs are shown in Figure~\ref{fig:NVIDIA} for both the \pz and \pr benchmarks. In both cases, the native CUDA implementation is used as the reference implementation, and the measurement time includes the kernel execution only. 
Several different compilers are used for the different implementations: the CUDA, Alpaka, and Kokkos implementations are compiled with the nvcc compiler; OpenMP and OpenACC are compiled with the OpenARC~\cite{openarc_hpdc14} compiler; the \stdpar versions are compiled with nvc++; and for SYCL, the intel/llvm-sycl repository is used to compile dpc++ tool chains to be compatible with NVIDIA/AMD GPUs.
For evaluation, whenever possible, we use the same launch parameters, including number of blocks and number of threads per block.  
In the OpenACC and OpenMP versions, different compilers varied in how they enforced the user-specified configurations (see Section~\ref{sec:compilers}).
The launch parameters need to be manually specified for Alpaka and Kokkos, otherwise the libraries choose suboptimal values and the performance is about 30\% worse, taking those versions even further from the native CUDA version. Setting the number of registers per thread is another approximately 10\% effect. These parameters cannot be manually specified in the \stdpar implementation. In the SYCL implementations, we specified the execution volume manually but relied on the defaults for the other launch parameters.

\setcounter{figure}{2}
\setcounter{subfigure}{0}
\begin{subfigure}
\setcounter{figure}{2}
\setcounter{subfigure}{0}
    \centering
    \begin{minipage}[b]{0.5\textwidth}
        \includegraphics[width=\linewidth]{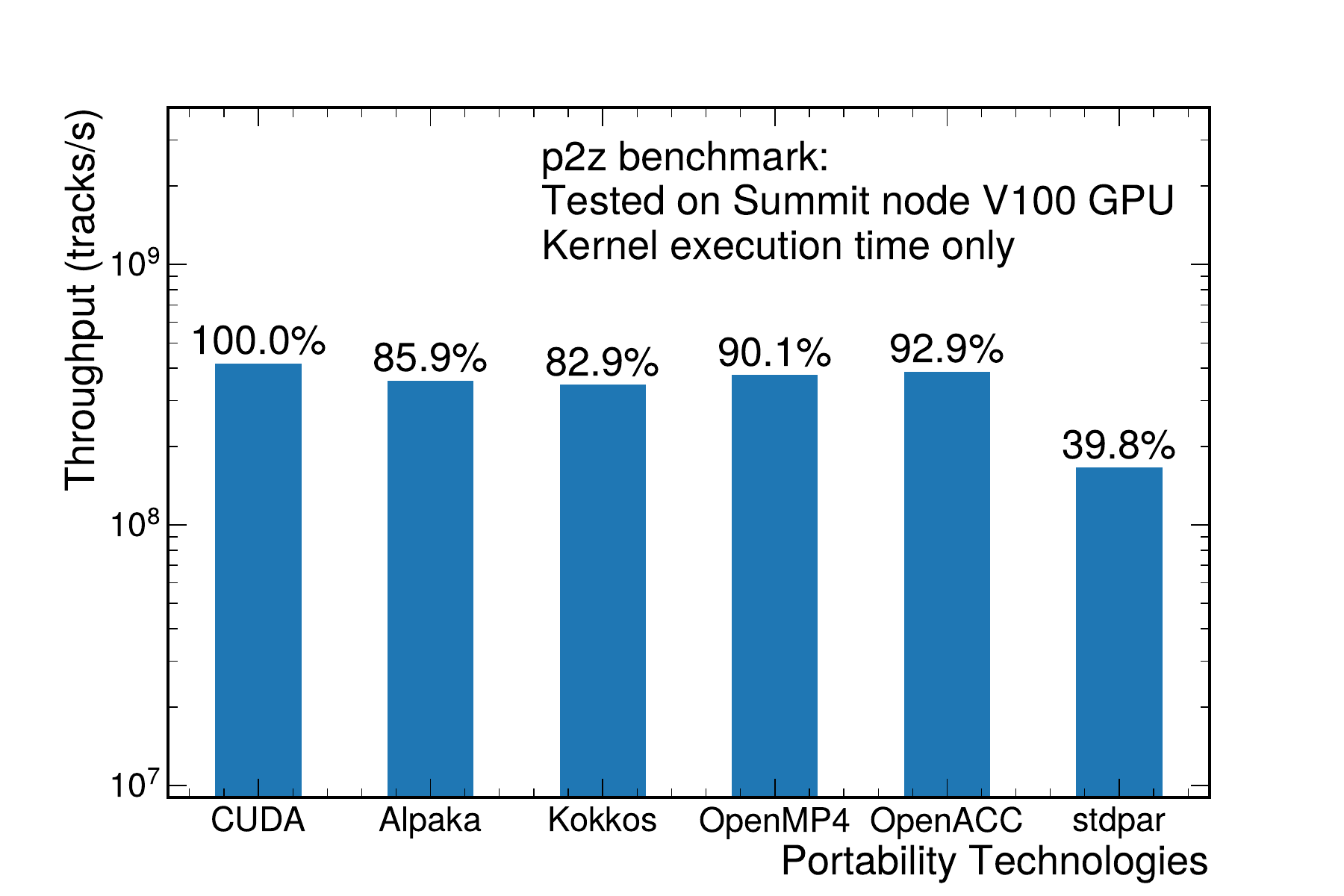}
    \end{minipage}  
   
\setcounter{figure}{2}
\setcounter{subfigure}{1}
    \begin{minipage}[b]{0.5\textwidth}
        \includegraphics[width=\linewidth]{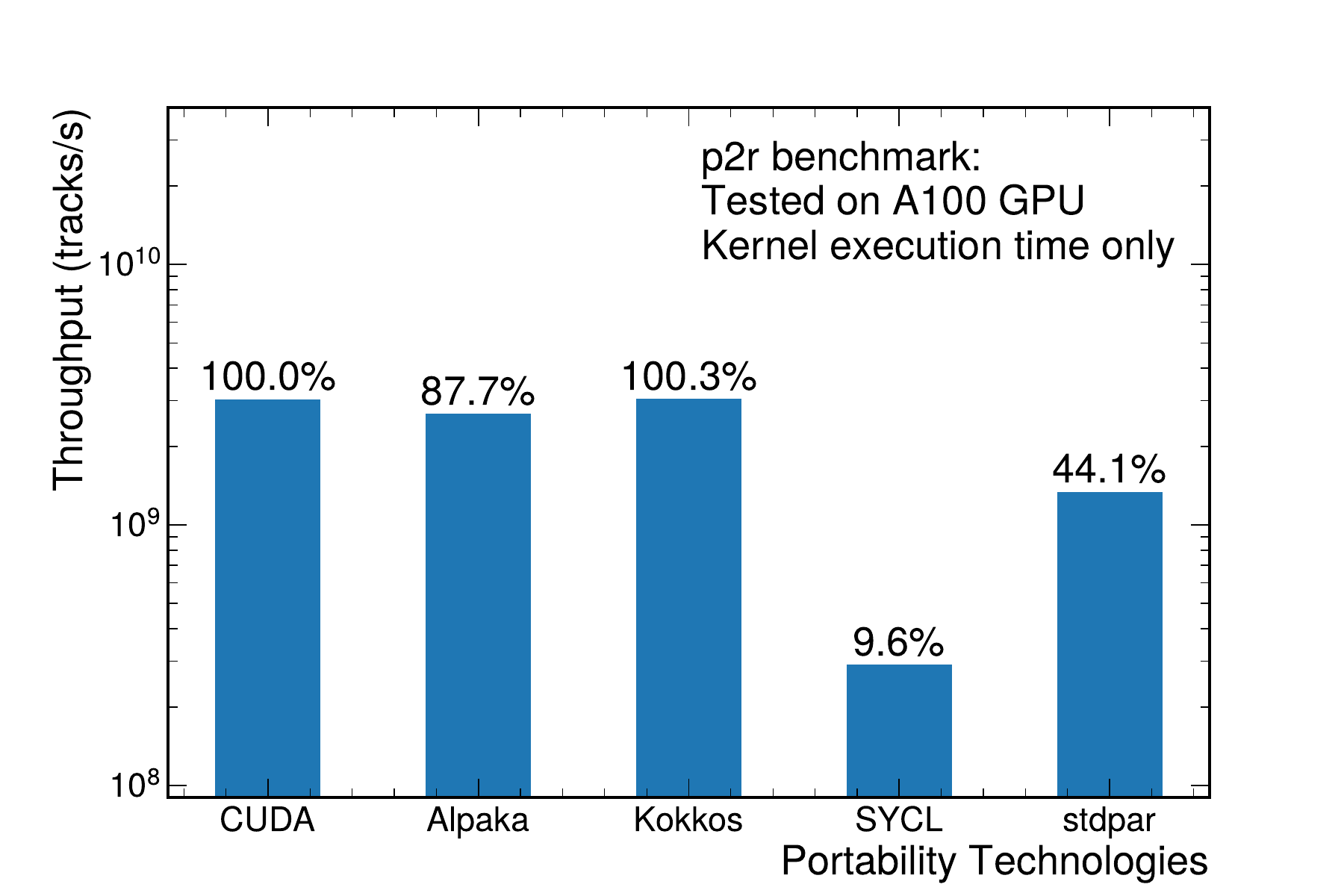}
    \end{minipage}

\setcounter{figure}{2}
\setcounter{subfigure}{-1}
    \caption{Throughput measurements for the \pz (top) and \pr (bottom) benchmarks on NVIDIA GPUs. The \pz measurements were performed on a V100 GPU and the \pr measurements were performed on an A100 GPU. In both cases, only the kernel execution time was considered in the throughput.}
    \label{fig:NVIDIA}
\end{subfigure}

In general, most of the different portability solutions managed to produce close-to-native performance. The \stdpar implementation did not perform well for either benchmark, mainly because the \stdpar implemented in the nvc++ compiler relies on CUDA Unified Memory for all data movements between CPU and GPU memory, which fetches data needed by the GPU kernel on demand, exposing memory transfer overheads to the kernel execution.
CUDA Unified Memory provides APIs to prefetch or migrate data between CPU and GPU memory to hide or reduce the memory transfer overheads, but the current \stdpar does not include such functionalities. 
In order to mitigate the effects of data transfers 
while measuring the kernel execution time, 
we introduced implicit prefetching using the parallel \texttt{std::copy} algorithm.
For the \pr benchmark, the worst performing version is the SYCL implementation. Detailed profiling using NVIDIA \texttt{NSight Compute} shows significant branching when using SYCL, but preliminary investigations have not revealed an obvious explanation for the branching. 

For the \pz benchmark, most of the different implementations achieved similar performance except for the \stdpar version.
While similar to other \pz implementations, the relative performance of \pz Kokkos version is lower than that of the \pr Kokkos version, when compared to their corresponding CUDA versions.
Detailed profiling shows that the Kokkos version of \pz uses more registers than the CUDA version, while the Kokkos version of \pr uses a similar number of registers compared to the CUDA version.





\begin{figure}[h!]
\begin{center}
\includegraphics[width=0.5\linewidth]{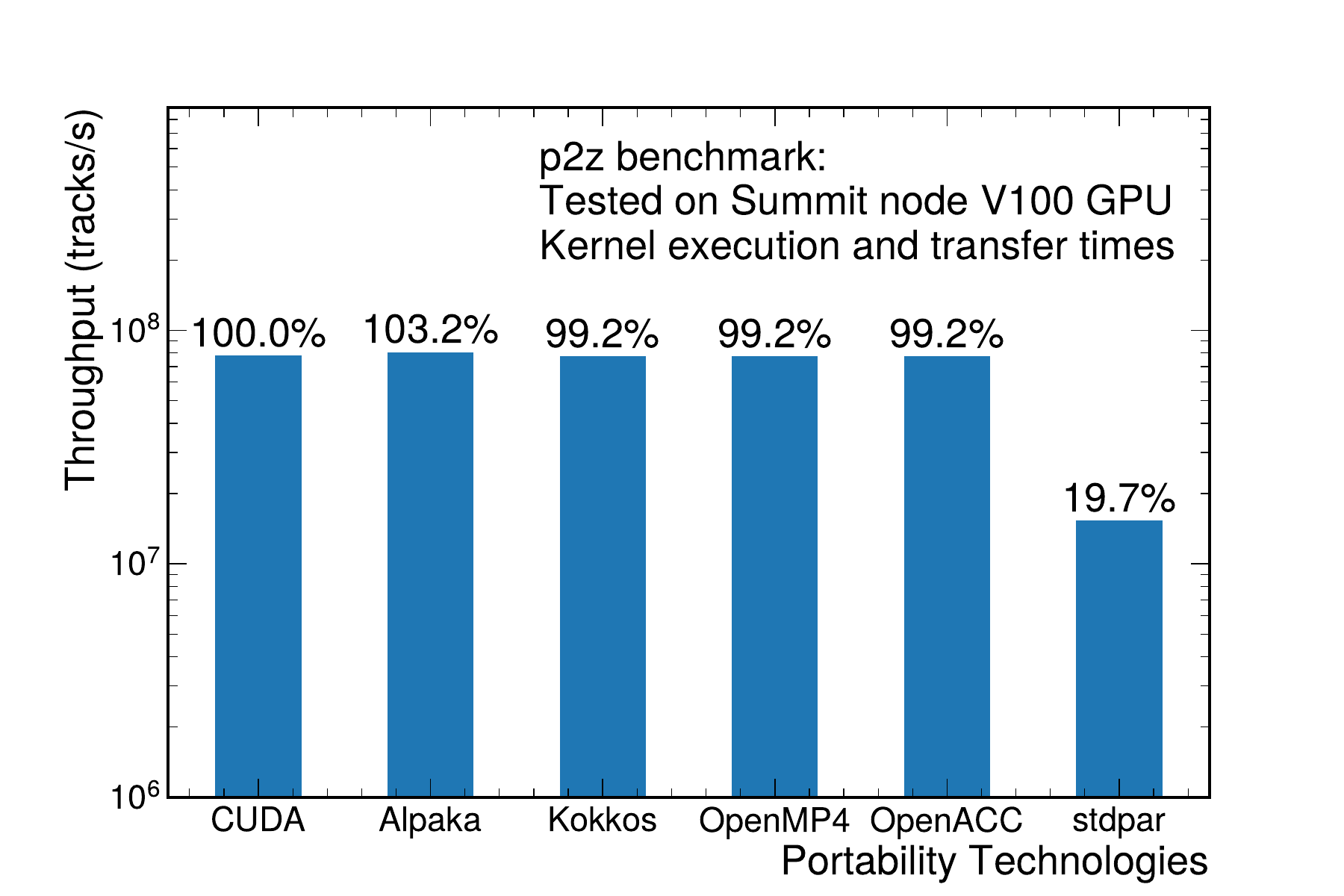}
\end{center}
\caption{Throughput measurement for the \pz benchmark on NVIDIA V100 GPUs, including kernel execution time as well as data transfer times.}
\label{fig:NVIDIA-kernel}
\end{figure}

Figure~\ref{fig:NVIDIA-kernel} shows the throughput of the \pz implementations on the NVIDIA V100 GPUs, this time including both the kernel execution time and the memory transfer times. The transfer times are generally 2 to 5 times larger then the kernel execution times, which means that much of the variability between implementations is concealed. With the exception of \stdpar, all implementations have close to identical performance because they provide memory management features to facilitate explicit data transfers, and in some cases memory prepinning as discussed in Sec. 5.1.2. 


\subsubsection{Compiler dependence}
\label{sec:compilers}

\begin{figure}[h!]
\begin{center}
\includegraphics[width=0.5\linewidth]{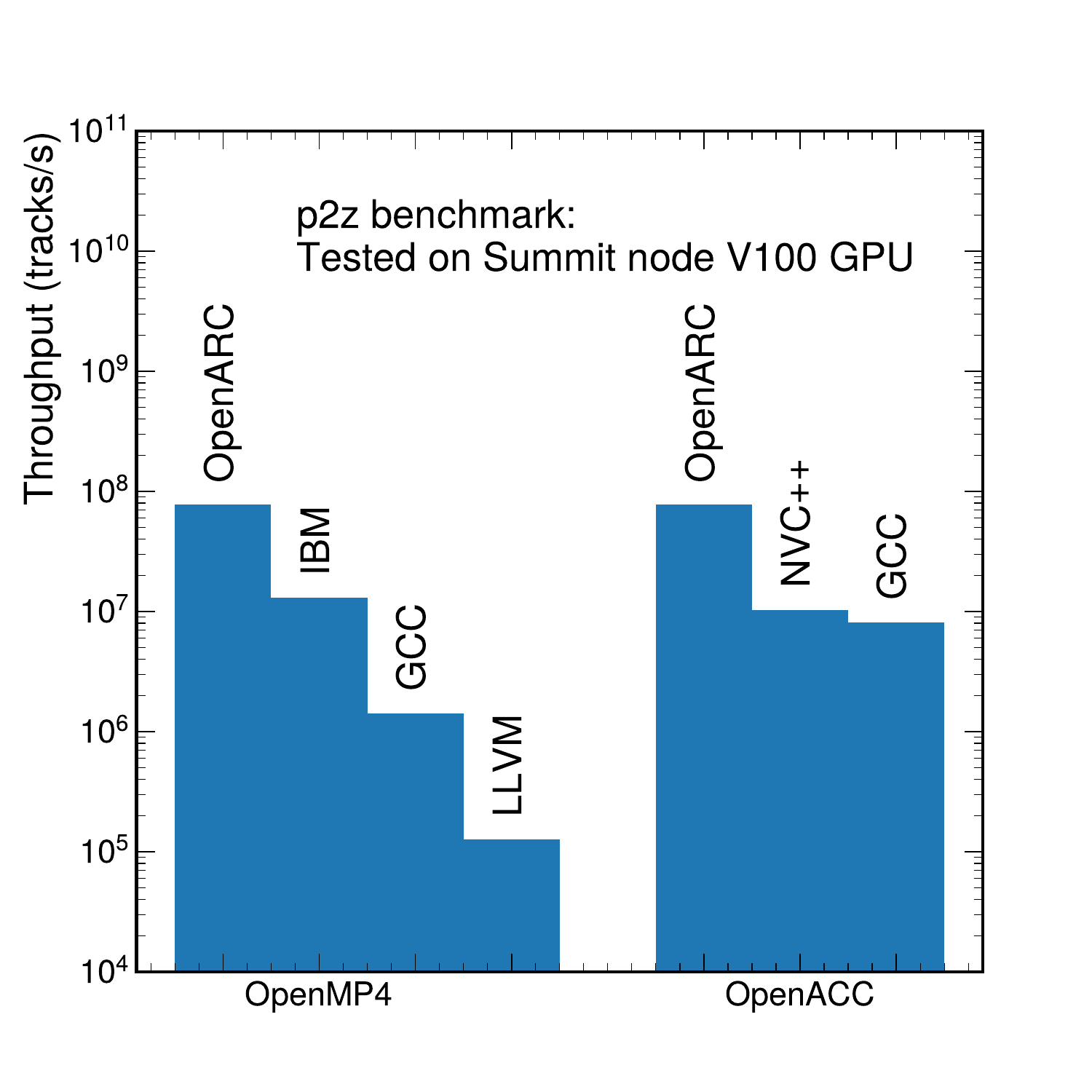}
\end{center}
\caption{ Comparison of the throughput performance on a V100 GPU using different compilers for the OpenMP (left) and OpenACC (right) versions of the \pz benchmark. Measurements include both kernel execution time and data transfer times. }
\label{fig:compilers}
\end{figure}

Different compilers can yield very different timing results, especially for the directive-based portability solutions. Figure~\ref{fig:compilers} shows the throughput performance for the OpenACC and OpenMP \pz versions on NVIDIA V100 GPUs, including both kernel execution time and data transfer times. 
The left bars in Figure~\ref{fig:compilers} show that the OpenARC-compiled OpenMP version performs better than the versions compiled with llvm, gcc, or IBM. 
Detailed profiling shows that the llvm, gcc, and IBM compiled versions use different launch parameters (the number of threads in a thread block and the number of thread
blocks) than those specified in the OpenMP program, while the OpenARC-compiled
version literally follows the user-specified configuration. Different launch parameters in the llvm/gcc/IBM compiled versions adversely affected the concurrency.
The OpenMP version of \pz allocates temporary user data in the team-private memory: the OpenARC-compiled version allocates the team-private data in the CUDA shared memory, but the llvm/gcc/IBM compiled versions partially use the CUDA shared memory, which incurs more device global memory accesses than the OpenARC-generated version. 
Lower concurrency and more global memory accesses seem to be the main reasons for the lower performance of the llvm/gcc/IBM compiled OpenMP versions.

In the OpenACC version of \pz (the right bars in Figure~\ref{fig:compilers}), both the nvc++ and OpenARC compiled versions achieved similar kernel computation times, but the OpenARC-compiled version had better memory transfer performance than the version compiled with nvc++. 
OpenARC literally translates the OpenACC data clauses into corresponding memory transfer APIs (one transfer API call per list item in a memory transfer data clause), but the nvc++ compiler automatically splits memory transfers into multiple, small asynchronous memory transfer calls.
Splitting memory transfers may expose more communication and computation overlapping opportunities, but in the \pz case, too many small asynchronous memory transfers in the nvc++ version perform worse than the simple memory transfers implemented in OpenARC, which is the main reason for the better performance of the OpenARC version than the nvc++ version.
Like OpenARC, gcc also generates one memory transfer API call per list item in a memory transfer data clause, but the gcc version achieves lower memory transfer throughput than the OpenARC version, which may be caused by the host memory pre-pinning optimization, which is supported by OpenARC but not by gcc, as shown in the next section.

\subsubsection{Effect of memory prepinning}




For GPU versions, implementing host memory pinning was found to greatly improve the performance, which is shown in Figure~\ref{fig:pinning} for the Kokkos and OpenACC \pz versions, including both kernel execution time and data transfer times. 
In the system with NVIDIA GPUs, prepinning the host memory enables direct-memory access (DMA) transfers, which achieve better memory transfer bandwidth than non-DMA transfers.

 We compared the performance with and without explicit prepinning for three OpenACC versions (Fig.~\ref{fig:pinning}, left).
 The first uses batched shared memory and synchronous transfers. The second implementation uses batched shared memory and asynchronous transfers, and the final version uses thread-private data on local memory with asynchronous transfers. Like the CUDA and HIP implementations (see Section~\ref{sec:cudaandhip}), the results show that keeping the intermediate results in the local register (thread-private data version) performs better than using the shared memory.
When host memory prepinning is on, all of the host data appearing in OpenACC clauses are prepinned.
In all cases, prepinning is observed to improve the performance.
The asynchronous versions have on-demand host-memory pre-pinning, which means the host memory is prepinned before each asynchronous memory transfer if it is not already prepinned. Therefore, a smaller impact from explicit prepinning is observed than for the synchronous transfer version. 
The performance improvement from the host memory prepinning is more pronounced in the Kokkos results (Fig.~\ref{fig:pinning}, right), because the internal implementation of the Kokkos library requires additional memory transfers, such as copying the functor objects, while the OpenACC/OpenMP versions do not.

\setcounter{figure}{5}
\setcounter{subfigure}{0}
\begin{subfigure}
\setcounter{figure}{5}
\setcounter{subfigure}{0}
    \centering
    \begin{minipage}[b]{0.5\textwidth}
        \includegraphics[width=\textwidth]{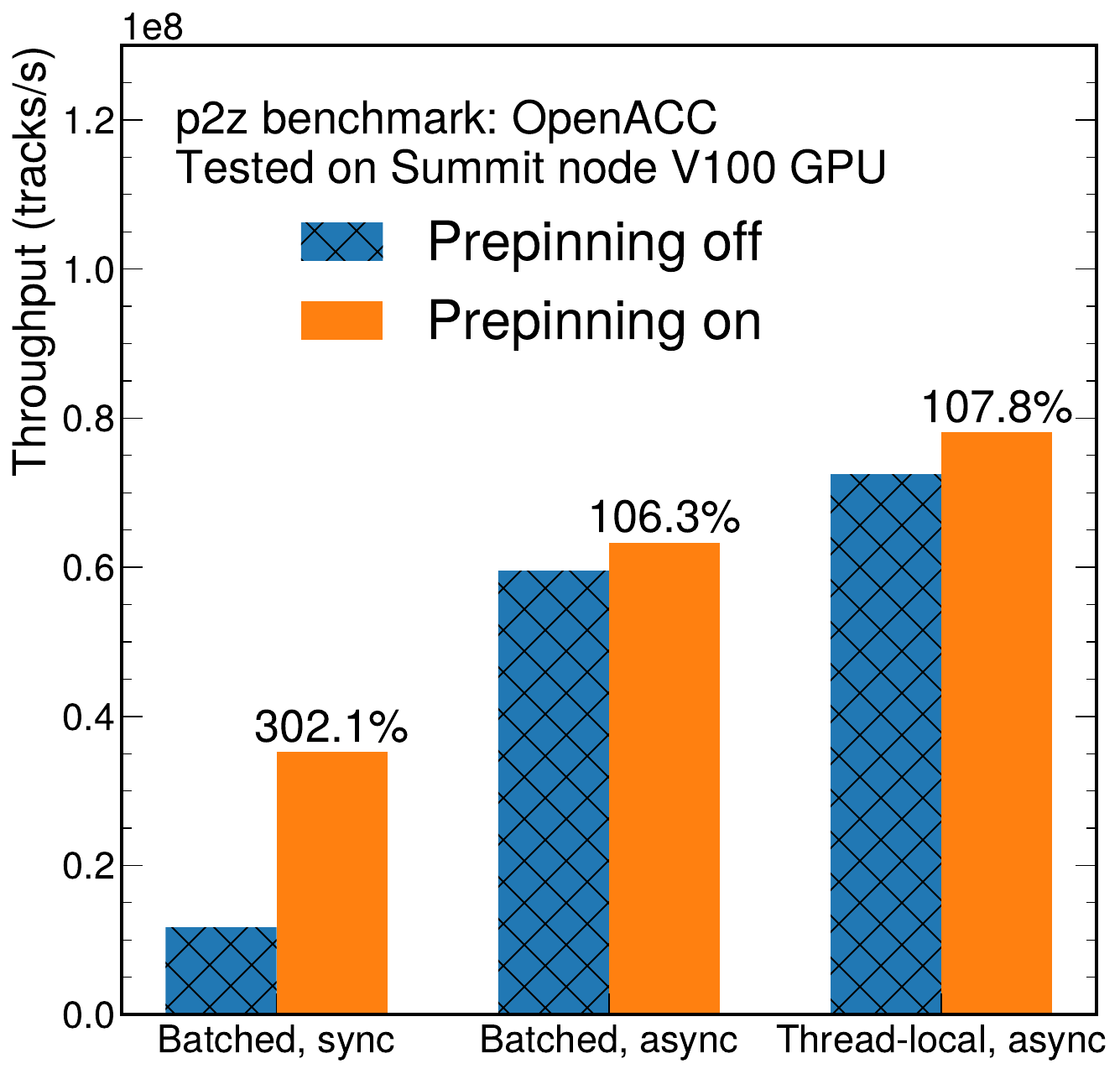}
    \end{minipage}  
   
\setcounter{figure}{5}
\setcounter{subfigure}{1}
    \begin{minipage}[b]{0.5\textwidth}
        \includegraphics[width=\textwidth]{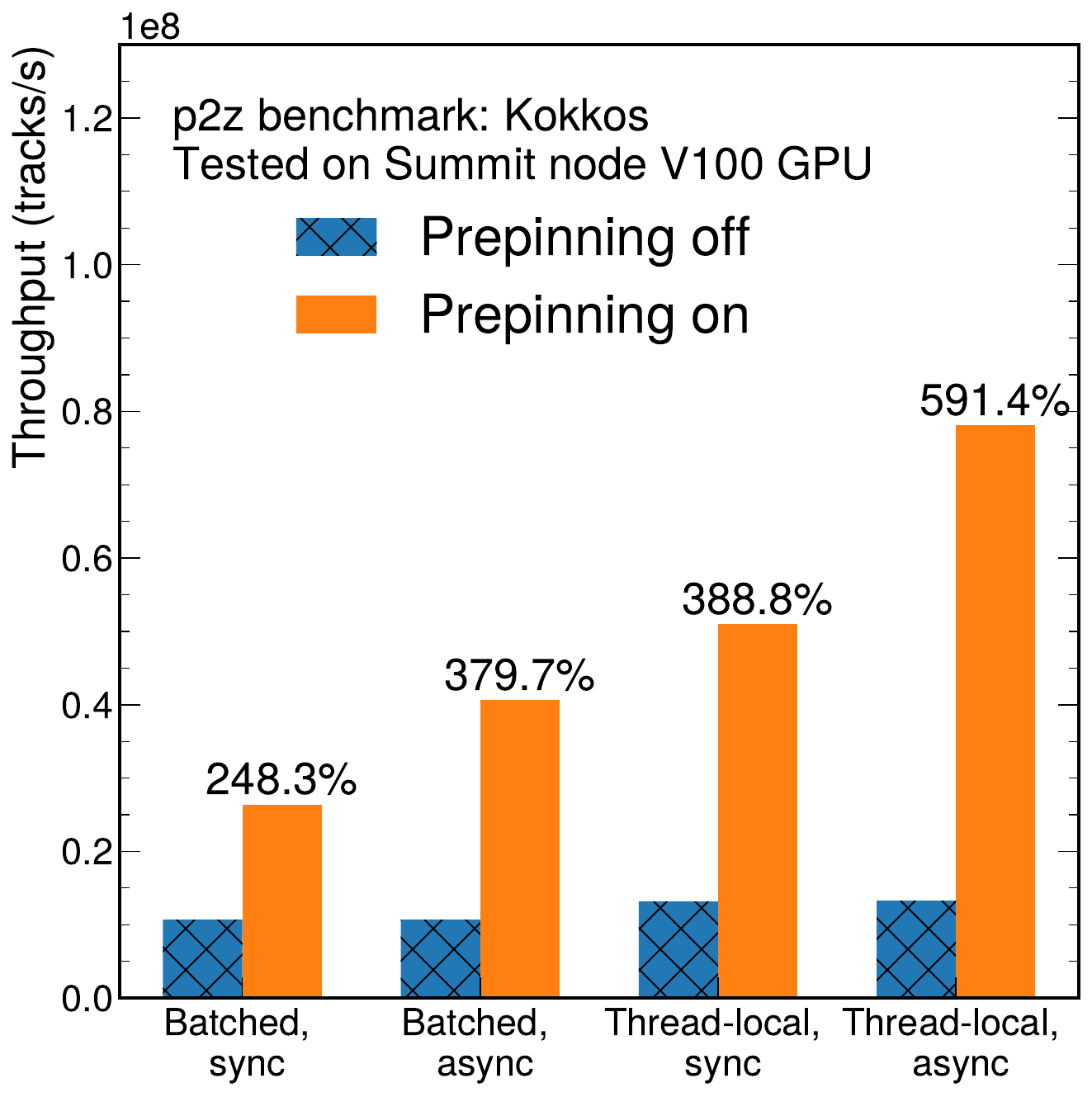}
    \end{minipage}

\setcounter{figure}{5}
\setcounter{subfigure}{-1}
    \caption{Throughput comparison on a V100 GPU showing the effect of turning memory prepinning on or off for different OpenACC (top) and Kokkos implementations (bottom) of the \pz benchmark: batch shared memory with synchronous or asynchronous transfers, and thread-private data on local memory with synchronous or asynchronous transfers. Measurements include both kernel execution time and data transfer times. The percentages in the figure refer to the throughput with memory prepinning on compared to the corresponding non-prepinned version.  }
\label{fig:pinning}
\end{subfigure}


\subsection{AMD and Intel GPU results} 
\label{sec:AMD_Intel}

Support for other GPU architectures is in general less mature than the support for NVIDIA GPUs, although this is an area of rapid expansion. We explored the preliminary performance of the \pr benchmark on both AMD and Intel GPUs, and the kernel-only throughput measurements are shown in Figure~\ref{fig:AMD_Intel}. 
This is an out-of-the-box comparison of each tool's portability; no dedicated effort was made to optimize for AMD or Intel GPU architectures. 
For example, the HIP implementation for the AMD GPU is a carbon copy of the native CUDA version.
For both Alpaka and Kokkos, switching backends is relatively seamless and does not require any code changes.

The results on the AMD GPU are shown in the left plot of Figure~\ref{fig:AMD_Intel}. The AMD GPU tests were performed on the JLSE testbed, which includes two AMD EPYC 7543 32c (Milan) CPUs and four AMD MI100 32GB GPUs. Only one GPU was used to perform the measurements.
Both Kokkos and Alpaka include HIP backends which achieve reasonable performance: Alpaka actually outperforms the HIP version that was ported from CUDA, and Kokkos is within a factor of about 2.  The same launch parameters as the CUDA implementation are used for the native HIP, Alpaka:HIP, and Kokkos:HIP measurements. 

The throughput measurements on an Intel A770 GPU are shown in the right plot of Figure~\ref{fig:AMD_Intel}. Since the A770 is not an HPC-class GPU, all calculations were converted to single-precision operations. Relying on double-precision emulation results in performance that is 3 to 30 times slower, depending on the implementation. 
The SYCL backend for Alpaka has only experimental support (introduced in v0.9.0) and was not tested here.
The Kokkos SYCL backend is still under active development: we observed a factor of 2 improvement in the throughput of the Kokkos:SYCL implementation when we updated from Kokkos 3.6.1 to Kokkos 4.1.0. 

\setcounter{figure}{6}
\setcounter{subfigure}{0}
\begin{subfigure}
\setcounter{figure}{6}
\setcounter{subfigure}{0}
    \centering
    \begin{minipage}[b]{0.5\textwidth}
        \includegraphics[width=\textwidth]{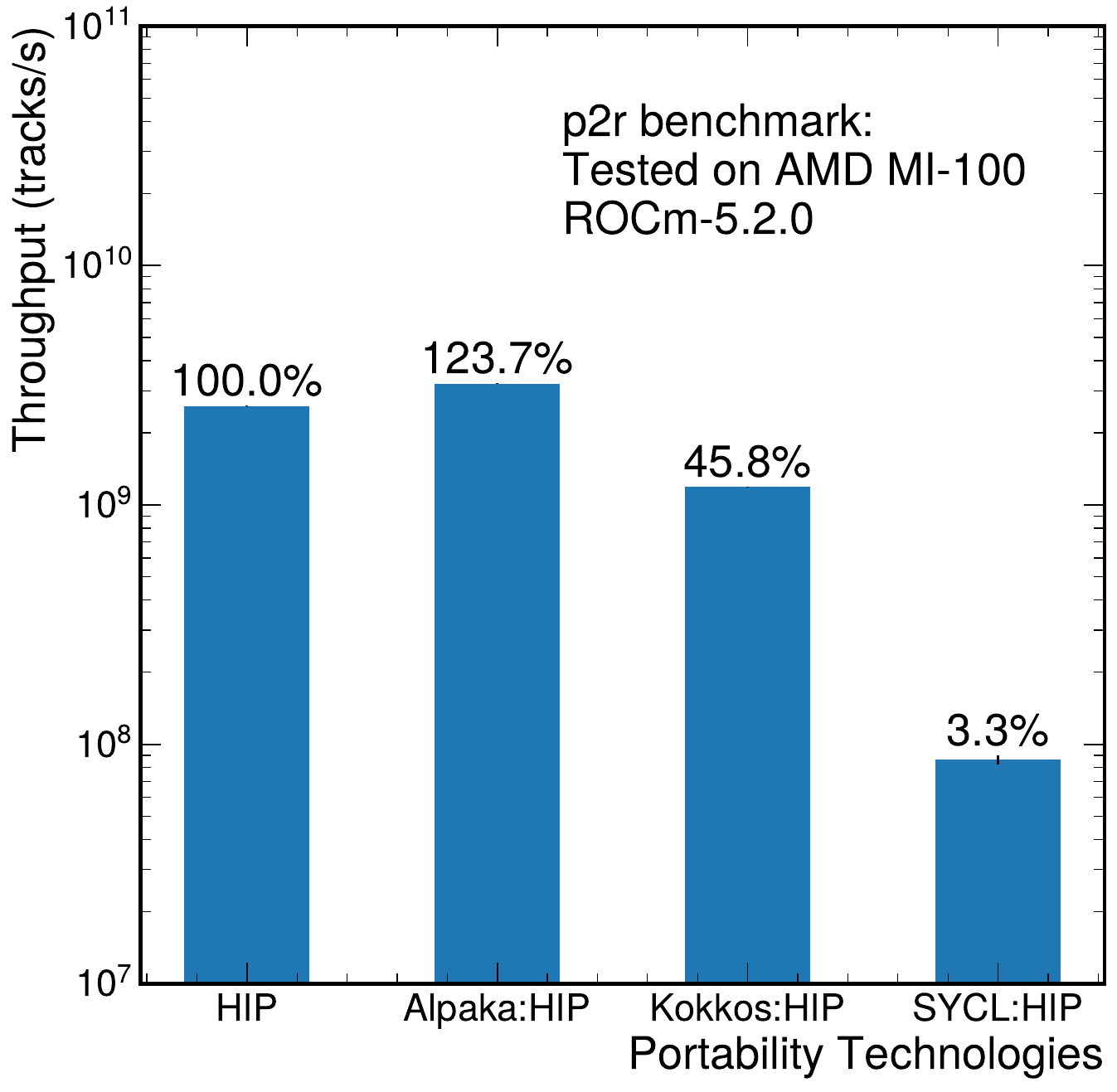}
    \end{minipage}  
   
\setcounter{figure}{6}
\setcounter{subfigure}{1}
    \begin{minipage}[b]{0.5\textwidth}
        \includegraphics[width=\textwidth]{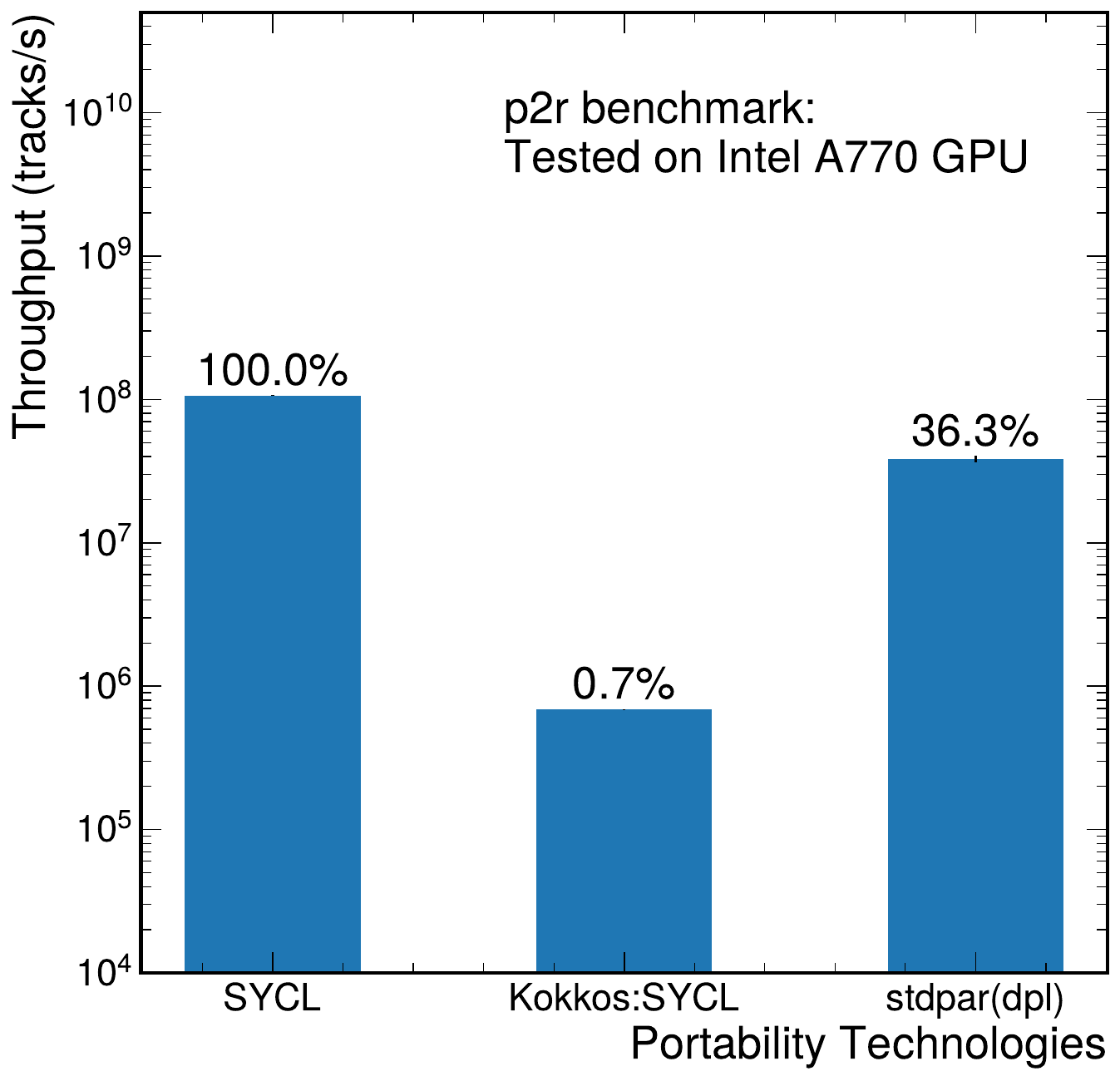}
    \end{minipage}

\setcounter{figure}{6}
\setcounter{subfigure}{-1}
    \caption{Throughput measurements for the \pr benchmark on an AMD MI-100 GPU (top) and on an Intel A770 GPU (bottom). The performance of each implementation is compared to the performance of the native version (HIP for the AMD GPU and SYCL for the Intel GPU). Only the kernel execution time was included in the measurements. }
    \label{fig:AMD_Intel}
\end{subfigure}


\subsection{CPU results}
\label{sec:CPU}

The original \mkFit application is parallelized using the Threading Building Blocks (TBB) library from Intel, so we also used TBB as the reference native implementation on the CPU. This implementation is  multithreaded and vectorized. It is worth noting that the original version was initially developed based on the ``classic'' Intel C++ Compiler (icc version 19), which led to improved vectorization performance compared to more recent compilers, resulting in \pz execution times that were approximately $2.7$x faster. However, since this version is not supported anymore, we choose not to include it in our main results. Figure~\ref{fig:CPU} shows the throughput performance of the \pz and \pr benchmarks on a two-socket system equipped with Intel Xeon Gold 6248 CPUs. 
All implementations were compiled with gcc except \stdpar, which was compiled with nvc++, and SYCL, which was compiled with dpcpp. 
Nonstandard options such as  \texttt{-ffast-math} are not included in the compilation since, while they may help vectorizing loops including trigonometric functions, they do not guarantee numerical reproducibility. 

Using the portability layers, we are able to achieve throughput equal to or better than 70\% of the original, native performance on the CPU for most implementations. The Alpaka implementation of the \pz benchmark actually outperformed the TBB reference implementation. 
The reason for the better performance of \pz Alpaka is not currently known, and the results might be platform dependent.
Note that the \pz Alpaka implementation uses the OpenMP execution backend, while the \pr Alpaka implementation relies on the TBB execution backend.  
The \pr SYCL implementation only achieves 27\% of the reference implementation, but unlike many of the other implementations, SYCL is a language extension and depends heavily on compiler optimizations rather than a performance-tuned library.
When optimizing these implementations, considerations included deciding on an optimal data layout that works for both CPUs and GPUs and ensuring that the loops are properly vectorized when run on the CPU.

\setcounter{figure}{7}
\setcounter{subfigure}{0}
\begin{subfigure}
\setcounter{figure}{7}
\setcounter{subfigure}{0}
    \centering
    \begin{minipage}[b]{0.5\textwidth}
        \includegraphics[width=\textwidth]{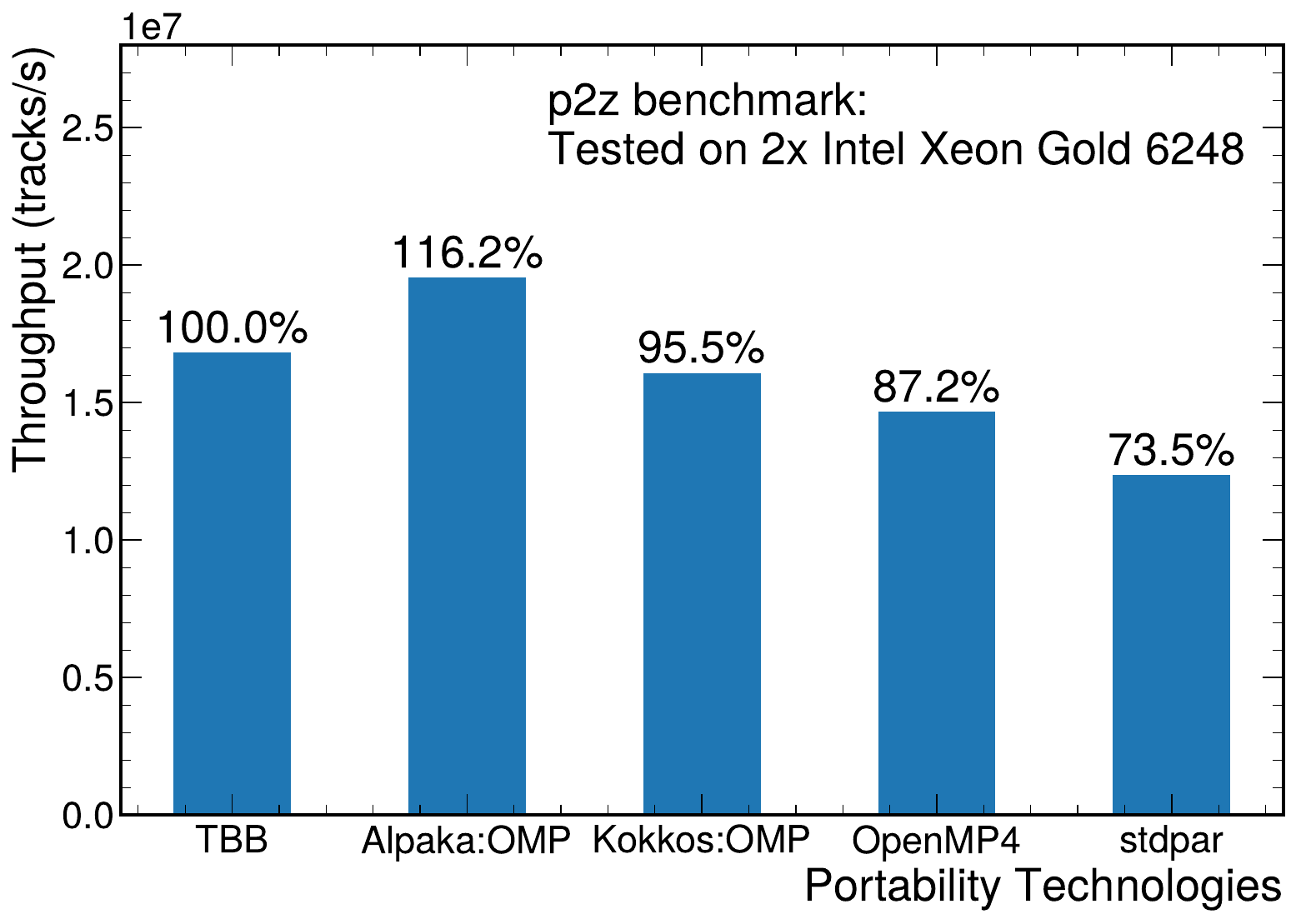}
    \end{minipage}  
   
\setcounter{figure}{7}
\setcounter{subfigure}{1}
    \begin{minipage}[b]{0.5\textwidth}
        \includegraphics[width=\textwidth]{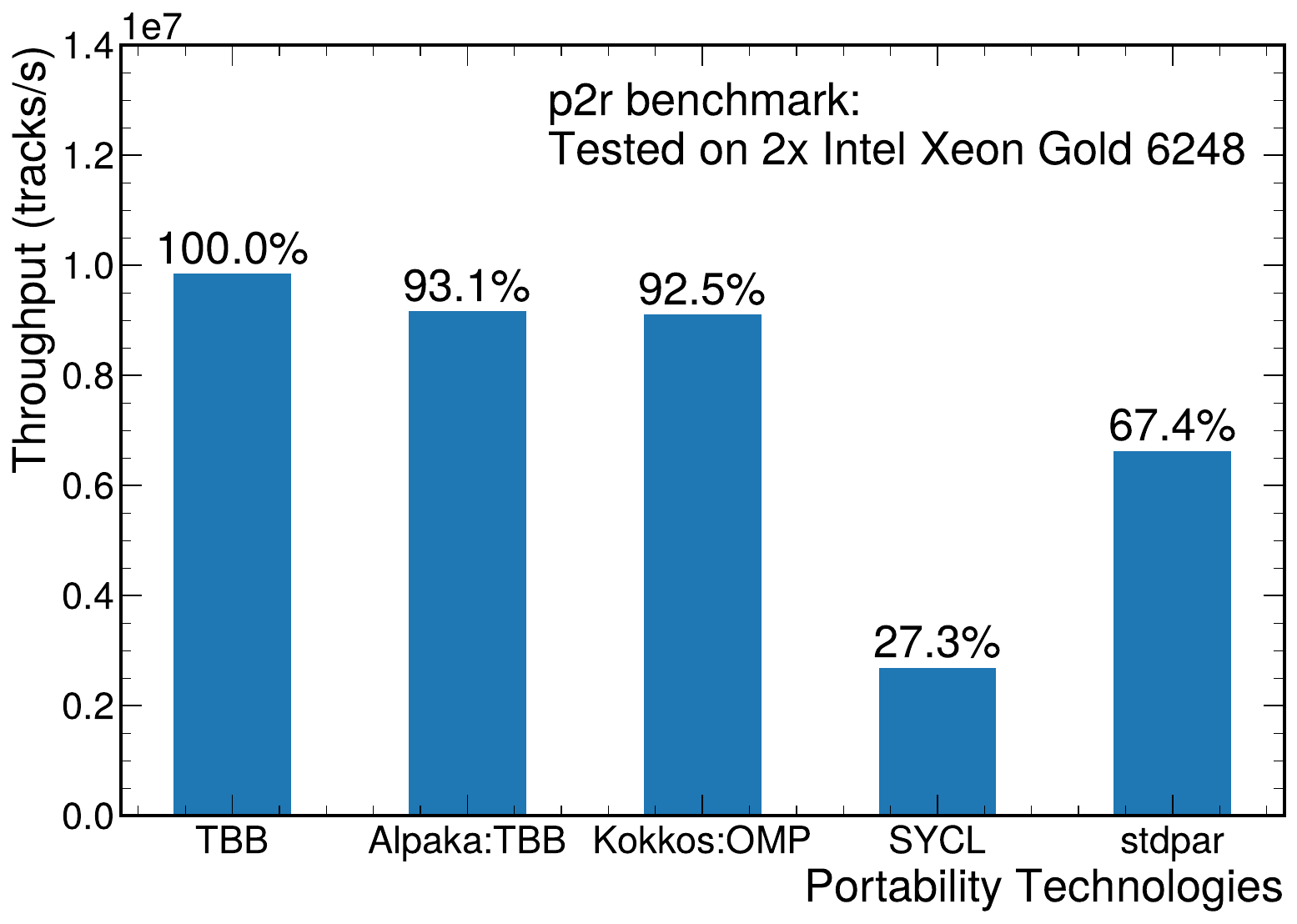}
    \end{minipage}

\setcounter{figure}{7}
\setcounter{subfigure}{-1}
    \caption{Throughput measurements for \pz (top) and \pr benchmarks (bottom) on a two-socket system equipped with Intel Xeon Gold 6248 CPUs. The performance of each implementation is compared to the performance of the native TBB version.  }
    \label{fig:CPU}
\end{subfigure}

\section{Summary}

\label{sec:summary}

In the project described in this paper, we ported two benchmark applications for charged-particle track reconstruction using state-of-the-art portability tools and compared the results to the native implementations. These benchmarks could form the backbone of a realistic track fitting algorithm. We have tested our ports on Intel CPUs and on GPUs from different vendors. 
In developing and testing these benchmarks, we found that the performance can vary significantly depending on the details of the implementation. Achieving optimal performance was not easy, even for relatively simple applications such as these. Each implementation took several iterations of profiling and development to achieve the results shown here. Furthermore, the steps that were taken to improve performance on one type of accelerator (NVIDIA GPUs, for example) did not necessarily translate into analogous gains on other types of GPUs or CPUs. 

Several factors were found to have large effects on the final performance. We found that optimizing the memory layout and enabling explicit memory prepinning (in the case of NVIDIA GPUs) led to big improvements in the performance of each implementation, up to a factor of six speedup. 
The choice of compiler also changed the throughput performance on NVIDIA GPUs by an order of magnitude or more for OpenMP and OpenACC implementations. Because these compilers are undergoing very active development, regularly checking performance with the latest versions is important. The same is true for the tools themselves, especially with respect to support for AMD and Intel GPUs. For example, we observed a factor of two speedup on an Intel GPU when updating to a newer version of the Kokkos library.

In summary, we explored major portability solutions using two benchmark applications from HEP,
including implementations using Alpaka, Kokkos, SYCL, \stdpar, OpenMP, and OpenACC. Most of these solutions can give reasonable performance, on the same order of magnitude, on different architectures, but most cases required significant optimization. 
The ability to run algorithms on GPUs from different vendors will allow HEP experiments to take advantage of a variety of computing resources, including current, planned, and future HPCs.
 This paper demonstrates that while tools exist to effectively port existing CPU algorithms to GPUs, reaching the desired performance is not yet straightforward. Future development of these portability solutions and their application, or alternative methods, will be necessary for the successful operation and data analysis of these experiments.


\section*{Conflict of Interest Statement}

The authors declare that the research was conducted in the absence of any commercial or financial relationships that could be construed as a potential conflict of interest.

\section*{Author Contributions}


\section*{Funding}

We thank the Joint Laboratory for System Evaluation (JLSE) for providing the resources for the performance measurements used in this work.
This research used resources of the Oak Ridge Leadership Computing Facility at the 
Oak Ridge National Laboratory, which is supported by the Office of Science of the 
U.S. Department of Energy under Contract No. DE-AC05-00OR22725.

This material is based upon work by the RAPIDS Institute and the ``HEP Event Reconstruction with Cutting Edge Computing Architectures'' project, supported by the U.S. Department of Energy, Office of Science, Office of Advanced Scientific Computing Research and Office of High-Energy Physics, Scientific Discovery through Advanced Computing (SciDAC) program.

This work was supported by the National Science Foundation under Cooperative Agreements OAC-1836650 and PHY-2323298; 
by the U.S. Department of Energy, Office of Science, Office of High Energy Physics under Award Number 89243023SSC000116;  
and by the U.S. Department of Energy, Office of
Science, Office of High Energy Physics, High Energy Physics Center for Computational Excellence (HEP-CCE) at Fermi National Accelerator Laboratory under B\&R KA2401045.

This manuscript has been authored by UT-Battelle, LLC under Contract No. DE-AC05-00OR22725 with the U.S. Department of Energy and by Fermi Research Alliance, LLC under Contract No. DE-AC02-07CH11359 with the U.S. Department of Energy, Office of Science, Office of High Energy Physics.  The publisher, by accepting the article for publication, acknowledges that the U.S. Government retains a non-exclusive, paid up, irrevocable, world-wide license to publish or reproduce the published form of the manuscript, or allow others to do so, for U.S. Government purposes. The DOE will provide public access to these results in accordance with the DOE Public Access Plan (http://energy.gov/downloads/doe-public-access-plan).


\section*{Data Availability Statement}
The software developed and analyzed for this study can be found in the  \texttt{p2z-tests} repository at https://github.com/cerati/p2z-tests/releases/tag/v1.0 for the \pz mini app and in the \texttt{p2r-tests} repository at https://github.com/cerati/p2r-tests/releases/tag/v1.0 for the \pr mini app.  

\bibliographystyle{Frontiers-Vancouver} 
\bibliography{p2z-paper}






\end{document}